\documentclass[aps,prl,twocolumn,preprintnumbers,showpacs,superscriptaddress,nofootinbib,]{revtex4}
\usepackage[usenames,dvipsnames]{xcolor}
\usepackage{graphicx}
\usepackage{amsmath,amsfonts,amssymb}
\usepackage{citesort}
\usepackage{epsfig}
\usepackage[hyperindex]{hyperref} 
\usepackage{latexsym}
\usepackage{graphicx}
\usepackage{amsmath}
\usepackage{amsfonts}   
\usepackage{amssymb}    
\usepackage{float}
\usepackage{bm}
\usepackage{breakurl}

\def\met{\ensuremath{E_{\mathrm{T}}^{\mathrm{miss}}}}

\def\tanb{\tan\beta}

\def\ee{e^+e^-\to h^0Z}

\def\lsim{\raise0.3ex\hbox{$\;<$\kern-0.75em\raise-1.1ex\hbox{$\sim\;$}}}
\def\gsim{\raise0.3ex\hbox{$\;>$\kern-0.75em\raise-1.1ex\hbox{$\sim\;$}}}

\def    \be            {\begin{equation}}
\def    \ee            {\end{equation}}
\def    \bea           {\begin{eqnarray}}
\def    \eea           {\end{eqnarray}}

\def \ie{{\it i.e.}}
\def \eg{{\it e.g.}}

\def\tanb{\tan \beta}

\def\sw2{sin^2 \theta_w}

\def\a^tau{\alpha_{\tau}}

\def\beq{\begin{equation}}
\def\eeq{\end{equation}}
\def\beqa{\begin{eqnarray}}
\def\eeqa{\end{eqnarray}}

\newcommand{\vevs}{\textrm{VEVs}}
\newcommand{\tev}{\,\textrm{TeV}}
\newcommand{\gev}{\,\textrm{GeV}}

\newcommand{\cm}{\,\textrm{cm}}

\newcommand{\fb}{\,\textrm{fb}}

\newcommand{\neutralino}{{\tilde{\chi}^0}}

\newcommand{\newc}{\newcommand}
\newc\BR{BR}
\newc{\akappa}{A_{\kappa} }
\newc\deltagmtwo{\delta (g-2)_{\mu}} 
\newc\deltaamu{\Delta a_{\mu}}

\def\anti{\overline}

\def\wt{\widetilde}
\def\MET{E_T \hspace{-1.2em}/\;\:}
\def\rpv{\not\!\!{R_p}}
\newc{\haa}{BR\(h_1\to a_1 a_1\)}
\newc{\abb}{BR\(a_1\to b\anti{b}\)}
\newc{\hbb}{BR\(h_1\to b\anti{b}\)}
\newc{\abund}{\Omega h^2}
\newc\bsgamma{b\rightarrow s \gamma }
\newc\bxsgamma{\overline{B}\rightarrow X_{s}\gamma}
\newc\brbsgamma{\BR(\overline{B}\rightarrow X_s\gamma)}

\begin{document}

\title{
Probing the ``$\mu$ from $\nu$'' supersymmetric standard model with
    displaced multileptons from the decay of a Higgs boson at the LHC
}

\author{Pradipta~Ghosh}
\email{pradipta.ghosh@uam.es} 
\affiliation{Departamento de F\'{\i}sica Te\'{o}rica, 
Universidad Aut\'{o}noma de Madrid,
Cantoblanco, 28049 Madrid, Spain}
\affiliation{Instituto de F\'{\i}sica Te\'{o}rica UAM--CSIC, 
Campus de Cantoblanco UAM, 28049 Madrid, Spain}

\author{Daniel~E.~L\'opez-Fogliani}
\email{daniel.lopez@df.uba.ar}
\affiliation{Departamento de F\'{\i}sica, Universidad de Buenos 
Aires \& IFIBA-CONICET, 1428 Buenos Aires, Argentina}

\author{Vasiliki~A.~Mitsou}
\email{vasiliki.mitsou@ific.uv.es}
\affiliation{Instituto de F\'{\i}sica Corpuscular CSIC--UV, 
c/ Catedr\'atico Jos\'e Beltr\'an, 2, 46980 Paterna (Valencia), Spain}

\author{Carlos~Mu\~noz} 
\email{carlos.munnoz@uam.es}
\affiliation{Departamento de F\'{\i}sica Te\'{o}rica, 
Universidad Aut\'{o}noma de Madrid,
Cantoblanco, 28049 Madrid, Spain}
\affiliation{Instituto de F\'{\i}sica Te\'{o}rica UAM--CSIC, 
Campus de Cantoblanco UAM, 28049 Madrid, Spain}

\author{Roberto~Ruiz~de~Austri}
\affiliation{Instituto de F\'{\i}sica Corpuscular CSIC--UV, 
c/ Catedr\'atico Jos\'e Beltr\'an, 2, 46980 Paterna (Valencia), Spain}


\begin{abstract}
The ``$\mu$ from $\nu$'' supersymmetric standard model ($\mu\nu$SSM) 
cures the $\mu$-problem and concurrently reproduces measured neutrino
data by using a set of usual right-handed neutrino superfields. 
Recently, the LHC has revealed the first scalar boson which naturally
makes it tempting to test $\mu\nu$SSM in the light of this
new discovery. We show that this new scalar while decaying
to a pair of unstable long-lived neutralinos, can lead
to a distinct signal with non-prompt multileptons. With
concomitant collider analysis we show that this signal
provides an intriguing signature of the model, pronounced
with light neutralinos. Evidence of this signal is well
envisaged with sophisticated displaced vertex analysis,
which deserves experimental attention.
\end{abstract}

\preprint{FTUAM-12-112,~~~IFT-UAM/CSIC-12-104,~~~IFIC-12-78}


\maketitle


The $\mu\nu$SSM~\cite{MuNuSSM,MuNuSSM2} contains in the superpotential 
$W$, in addition to the Yukawa couplings for quarks and charged leptons of the minimal supersymmetric standard model (MSSM) \cite{mssm}, 
Yukawas for neutrinos and two additional type of terms involving the Higgs 
doublet superfields, $\hat H_u$ and $\hat H_d$, and the 
three right-handed neutrino superfields
$\hat \nu^c_i$~\cite{MuNuSSM,MuNuSSM2}:
{\small
\begin{align}\label{superpotential}
W  &= 
\ \epsilon_{ab} (
Y_{u_{ij}} \, \hat H_u^b\, \hat Q^a_i \, \hat u_j^c +
Y_{d_{ij}} \, \hat H_d^a\, \hat Q^b_i \, \hat d_j^c +
Y_{e_{ij}} \, \hat H_d^a\, \hat L^b_i \, \hat e_j^c 
\nonumber\\ 
&
+ Y_{\nu_{ij}} \, \hat H_u^b\, \hat L^a_i \, \hat \nu^c_j -   
\lambda_{i} \, \hat \nu^c_i\,\hat H_d^a \hat H_u^b)+
\frac{1}{3}
\kappa{_{ijk}} 
\hat \nu^c_i\hat \nu^c_j\hat \nu^c_k\ .
\end{align}}
The simultaneous presence of the last three terms 
in Eq.~(\ref{superpotential}) gives rise to explicit
breaking of $R$-parity ($R_p$).
With only dimensionless trilinear couplings in $W$, 
the electroweak (EW) scale arises through the soft supersymmetry 
(SUSY)-breaking terms in the scalar potential.
Thus all known particle physics phenomenology can be reproduced 
in the $\mu\nu$SSM with only one scale. Once the EW symmetry is spontaneously 
broken, the neutral scalars develop in general the following 
vacuum expectation values (VEVs): 
$\langle H_d^0 \rangle = v_d$, $\langle H_u^0 \rangle = 
v_u  $, $\langle \tilde \nu_i \rangle = \nu_i$, $ \langle \tilde \nu_i^c 
\rangle = \nu_i^c $.
An effective interaction $\mu\,\hat H_d \hat H_u$, with $\mu\equiv 
\lambda_i \nu_i^c$, is generated through the fifth term of Eq.~(\ref{superpotential}), 
solving the $\mu$-problem of the MSSM \cite{muproblem}
without introducing 
an extra singlet superfield as in the 
case of the next-to-MSSM (NMSSM) \cite{ana}. The sixth term 
in Eq.~(\ref{superpotential}) avoids the existence of a Goldstone boson.
It also generates EW-scale effective Majorana masses ($2\kappa \nu_i^c$) 
for right-handed neutrinos, which give rise to a TeV scale seesaw
with $Y_\nu\sim 10^{-6}$ (like $Y_e$), and
together with $R_p$ violation $(\rpv)$ are instrumental in reproducing the measured 
neutrino mass squared differences and mixing 
angles~\cite{MuNuSSM2, MuNuSSM SCPV, Ghosh2008,Ghosh2010} at the tree level.
This feature is unlike 
the bilinear $\rpv$ model \cite{Barbier:2004ez} where only one mass 
is generated at the tree level and loop corrections are necessary to generate at least a second mass and a PMNS mixing matrix compatible with experiments. In the bilinear model, the $\mu$-like problem \cite{Nilles} is also augmented with three bilinear terms.

In the $\mu\nu$SSM as a consequence of the $\rpv$,  all the neutral 
fermions (scalars) mix together and there are 10 neutralino (8 $CP$-even and 7
$CP$-odd) mass eigenstates.
Analyses of the $\mu\nu$SSM, with attention to the neutrino and 
LHC phenomenology have also been addressed in 
~\cite{MuNuSSM2,Ghosh2008,Ghosh2010,Hirsch2009,Ghosh2010cu,Porod,MuNuLHC}.
Other analyses concerning cosmology such as gravitino dark matter and electroweak baryogenesis can be found in~\cite{gravitino} and~\cite{baryogenesis}, respectively.

Thus the $\mu\nu$SSM is a well motivated SUSY model with
enriched phenomenology and notable signatures, which
definitely deserve rigorous analyses by the LHC collaboration.
However, although SUSY searches remain one of the primary
targets for the LHC, the discovery of a new
scalar boson with a mass around $125\gev$ by ATLAS~\cite{ATLAShiggs} 
and CMS~\cite{CMShiggs} collaborations has attracted
the attention of the community.
In spite of the observed decay rates of 
this particle compatible with those of the standard model (SM) Higgs boson, 
a departure from the SM predictions remains a possibility since new LHC data 
are being analysed. 
We present a dedicated collider analysis 
together with detector simulation of an intriguing 
signal in the $\mu\nu$SSM featuring 
non-prompt multileptons at the LHC, arising from the 
beyond SM decay of a  $125 \gev$ scalar into a pair 
of lightest neutralinos ($\widetilde \chi^0$).
Since $R_p$ is broken, each $\widetilde \chi^0$ decays into a 
scalar/pseudoscalar $(h/P)$ and a neutrino ($\nu$), 
with the $h/P$ further driven to decay into $\tau^+ \tau^-$, 
giving rise to a $4\tau$ final state. 
Because of the value of $Y_{\nu}$, $\rpv$ is small and 
$\widetilde \chi^0$ decay leads to a displaced vertex (DV). 
We investigate the situation
when the $\widetilde \chi^0$ decays inside the inner tracker and thereby
yielding clean detectable signatures.
Although $h/P \to b\bar{b}$ is dominant over a 
broad range of parameters, we stick to 
$2m_{\tau}\lesssim m_{h/P}\lesssim 2m_b$, which
allows us to demonstrate two possible signatures characterised 
by: (i) high lepton multiplicity; and (ii) charged tracks originating 
from DVs, which can be explored through distinct 
experimental approaches. 

It remains to elucidate, if a possible excess of four-lepton 
($e,\mu$ coming from $\tau$-decay) events is observed at the LHC, 
how to distinguish the $\mu\nu$SSM from other models. 
A dedicated measurement of the displaced charged tracks,
in the first hand, can reject all possible similar
final states with prompt leptons. 
In bilinear $\rpv$ models with minimal superfield content,   
$\widetilde\chi^0$ with mass below $m_W$ dominantly decays 
through $\ell W^*/\nu Z^*$ with $\ell=e,\mu,\tau$, while
decay length ($l_{DL}$) scales as $1/m_{\widetilde \chi^0}^4$.
In the $\mu\nu$SSM however, available lighter singlet
like $h/P$ states provide new two-body
decay modes, $\widetilde\chi^0 \to h/P +\nu$, which
can reduce $l_{DL}$ by orders of magnitude, as hinted in~\cite{Hirsch2009}. 
This feature manifests notably in the regime 
$m_{\widetilde \chi^0}<20~\gev$ where bilinear $\rpv$ models
predict $l_{DL}\sim 100$ m~\cite{Hirsch2009}, beyond
detector coverage and thus mimic known missing energy 
$(\MET)$ signature with conserved $R_p$.
On the other hand, models with trilinear $\rpv$ term, for 
example $\hat L_i \hat L_j \hat E^c_k$ 
\cite{Barbier:2004ez}, can produce moderate to large displaced 
vertex (1cm-3m) for $m_{\widetilde \chi^0}<20$ GeV.
Such a light $\widetilde\chi^0$ within minimal SUSY models,
will however, give rise to high branching ratio for 
SM-like Higgs $\to \widetilde\chi^0\widetilde\chi^0$ decay
mode. Since resulting final states contain visible
particles this scenario is tightly constrained from
experimental data. This problem is however ameliorated
in the $\mu\nu$SSM since the said branching fraction
can still remain suppressed due to predominant singlet
composition of $\widetilde \chi^0$. 

Wrapping up, moderately displaced 
($\gsim 1$ cm) yet detectable charged tracks 
from a light ${\widetilde \chi^0}$, as appear in the $\mu\nu$SSM, are hardly 
possible with other minimal SUSY models with or without $R_p$. 
Non-minimal SUSY models with or without $\rpv$ can however 
produce similar displaced final states. As an example, in the NMSSM
these states can appear when
a light next-to-lightest supersymmetric particle 
(NLSP) (produced directly or from Higgs decay) decays to a very light LSP
(to yield small $\MET~$) and a light scalar/pseudoscalar. 
This scenario in reality, however, is a hardly realistic
choice with experimental measurement of 
invisible $Z$-decay width and other LEP (and LHC) measurements.
On the contrary, non-minimal SUSY models with $\rpv$ 
produce irreducible impostor for this specific signal.
Nevertheless, there exists other unique decay modes, for
example multilepton final states from long Higgs-to-Higgs cascade \cite{MuNuLHC},
which can provide distinctive signal of the $\mu\nu$SSM because more Higgses are present.
Is is also worth remarking in this context that whereas the trilinear $\rpv$ terms in the $\mu\nu$SSM are useful for reproducing neutrino data at tree level and for solving the 
$\mu$-problem, trilinear $\rpv$ terms such as $\hat L_i \hat L_j \hat E^c_k$ do not
attempt to solve the $\mu$-problem and generate 
neutrino masses only through loops.

The noticeable footprint of the $\mu\nu$SSM relies on
the presence of light $\widetilde\chi^0$ and lighter $h,P$,
which are experimentally feasible when predominantly 
singlet in nature. Concealing complex flavour structure
and neglecting terms $\propto Y_\nu,\nu$ for smallness,
$m_{\widetilde\chi^0} \sim 2\kappa \nu^c$,
for $|2\kappa\nu^c|\ll|\mu|,|m_{\rm{gaugino}}|$, 
which is favoured with small $\kappa$. 
Also, in the limit of moderately small 
$\lambda~(\sim 0.1)$ singlet like $h,P$ get already decoupled
from the doublet sector and one 
gets from Ref.~\cite{MuNuSSM2},
$m^2_{h}\sim m^2_{\widetilde\chi^0} + m_{\widetilde\chi^0}A_\kappa/2$
and $m^2_{P}\sim - 3m_{\widetilde\chi^0} A_\kappa/2$. 
In the small $\lambda$ limit, $\nu^c\sim 1$ TeV is
also apparent as $\mu~(\sim 3\lambda \nu^c) \gsim 100$ 
GeV from chargino searches.
Thus, in the region of interest, that is 
$2m_{\tau}\lsim |m_{\widetilde\chi^0}|\lsim20$ GeV,
naively $10^{-3}\lesssim|\kappa| \lsim 10^{-2}$, and from the constraint
$2m_{\tau}\lesssim m_{P}\lesssim 2m_b$ one also obtains 
$0.4\lesssim|A_\kappa|\lesssim 30$~GeV.
Among other relevant parameters,
small tan $\beta$ seems useful to evade LEP constraints.
The aforesaid discussion is an artifact of the non minimal
nature of the $\mu\nu$SSM and thus could be related to well 
studied ~\cite{nmssm1,nmssm2} corners of the NMSSM parameter space
with a similar spectrum. Being precise, a light $P$ in the NMSSM is 
related to a solution of the little hierarchy problem
by using of small $A_\kappa$~\cite{nmssm2}.  
Light $\widetilde\chi^0$ and $h$, on the other hand,
are related to revival of an approximate Peccei-Quinn symmetry
of NMSSM in the limit of vanishing $\kappa$~\cite{nmssm1,nmssm2}.
Although, spectrum of the $\mu\nu$SSM is grossly enriched with three singlet 
$\hat \nu^c_i$ and $\rpv$, still presence of light $h,P$ 
and $\widetilde\chi^0$ in the model is well motivated with non-minimal nature.
This region of parameter space also contains a clear finger print of 
the $\mu\nu$SSM through detectable DVs for a very light $\widetilde\chi^0$.

In what follows, neutralino mass eigenstates are denoted by 
$\neutralino_{1,...,10}$, where $\neutralino_{i}$, with $i=1,2,3$,
coincide with the three left-handed neutrinos. 
$\neutralino_{4}$ has been identified as the lightest neutralino, 
with leading right-handed neutrino composition 
at the limit of small $\kappa$. 
$CP$-even ($CP$-odd) scalar mass eigenstates 
are denoted by $h_{1,...,8} (P_{1,...,7})$.
In the chosen benchmark, $h_4$
is basically decoupled from the right-handed sneutrinos 
$\tilde \nu_i^c$ and is the lightest doublet-like Higgs.
Three lightest scalar $h_{i}$ (pseudoscalar $P_i$) states
are the ones composed mainly of $\tilde \nu_i^c$.

As in Ref.~\cite{MuNuSSM2},
we eliminate eight soft masses 
in favor of the corresponding VEVs. We chose 
$\nu^c_i=780 \gev$, $\tanb \approx \frac{v_u}{v_d}=3.7$ and 
$v\approx\sqrt{v^2_u+v^2_d} \approx 174 \gev$. We assume gaugino mass unification 
at the GUT scale, and consider  $M_2=500 \gev$ at the EW scale.
We have fixed the following universal soft parameters at low energy:  
$m_{\tilde{e}^c}= m_{\tilde{u}^c}= m_{\tilde{d}^c}= m_{\tilde{Q}}=1 \tev$,  
$A_{\lambda}=990 \gev$, $A_{\kappa}= 5 \gev$, $A_e=A_d=-A_{\nu}=1 \tev$ 
and $A_u=2.4 \tev$. 
Chosen values for $m_{\tilde{Q}}$ and $A_u$ are crucial for
a sizable loop corrections to reach $m_{h_4}\sim 125~\gev$ with
selected $\lambda_i$.
At the limit of moderately small $\lambda$ (and small $\kappa$), 
the value of $A_\lambda$ is relevant for doublet-singlet mixing. 
We have checked that for the chosen values of the parameters, $A_\lambda$ 
can be varied in the range $980 \lsim A_\lambda \lsim 1040$~GeV 
without changing the studied signal significantly.
The remaining soft parameters can be arbitrarily changed without altering 
significantly the discussion presented here.

%

\begin{figure*}[t]
\centering
\epsfig{file=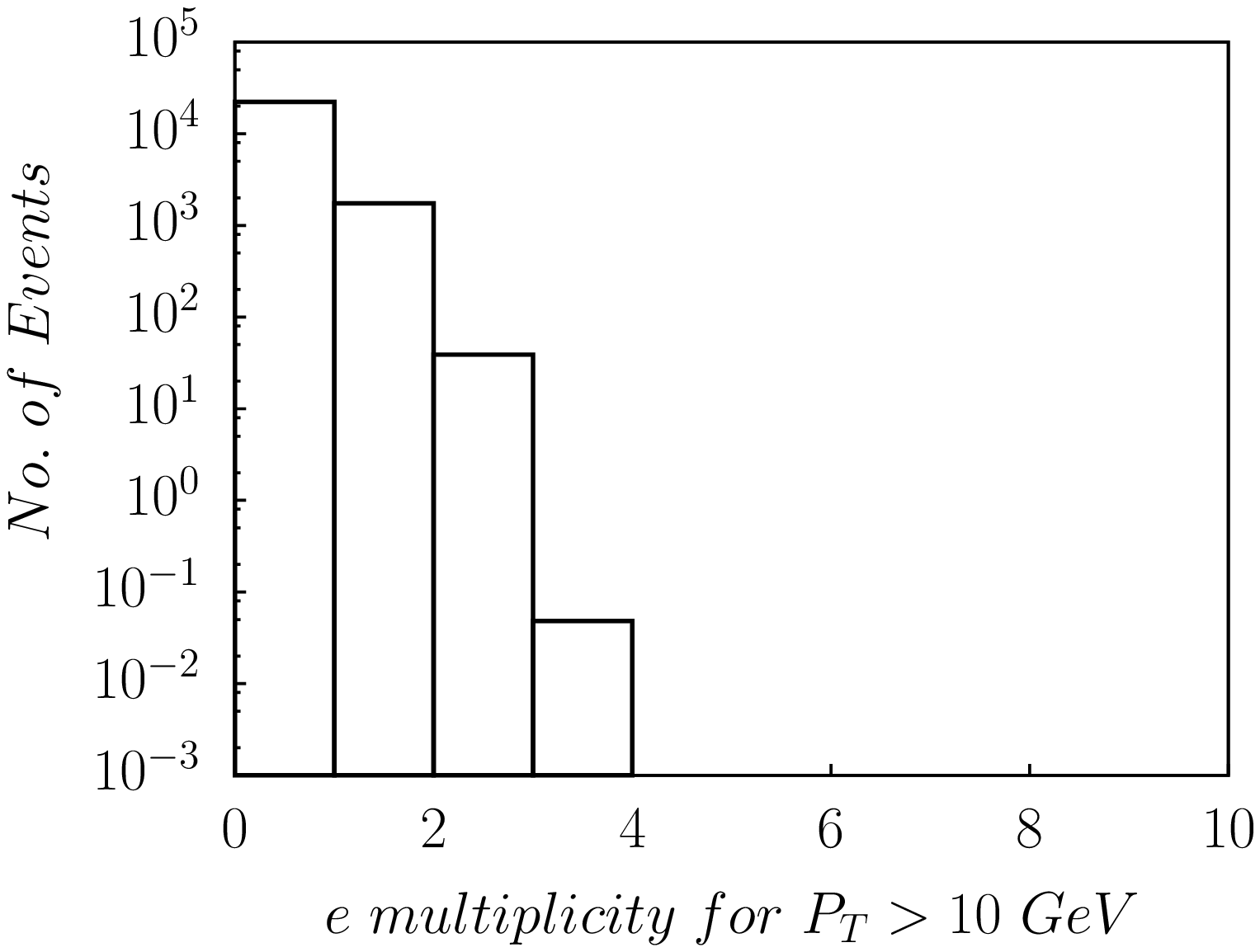,width=5.4cm,height=2.7cm}
\epsfig{file=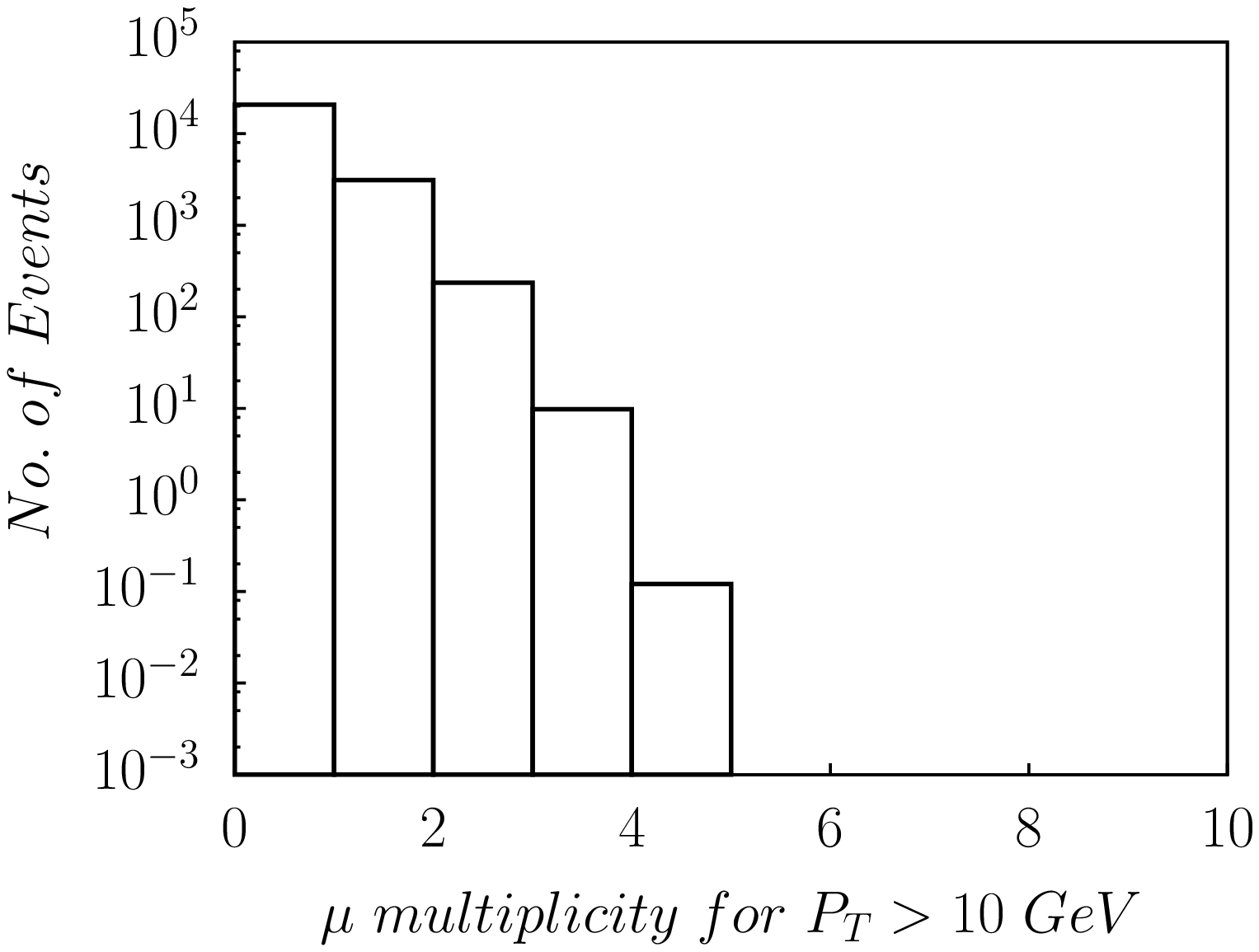,width=5.4cm,height=2.7cm}
\epsfig{file=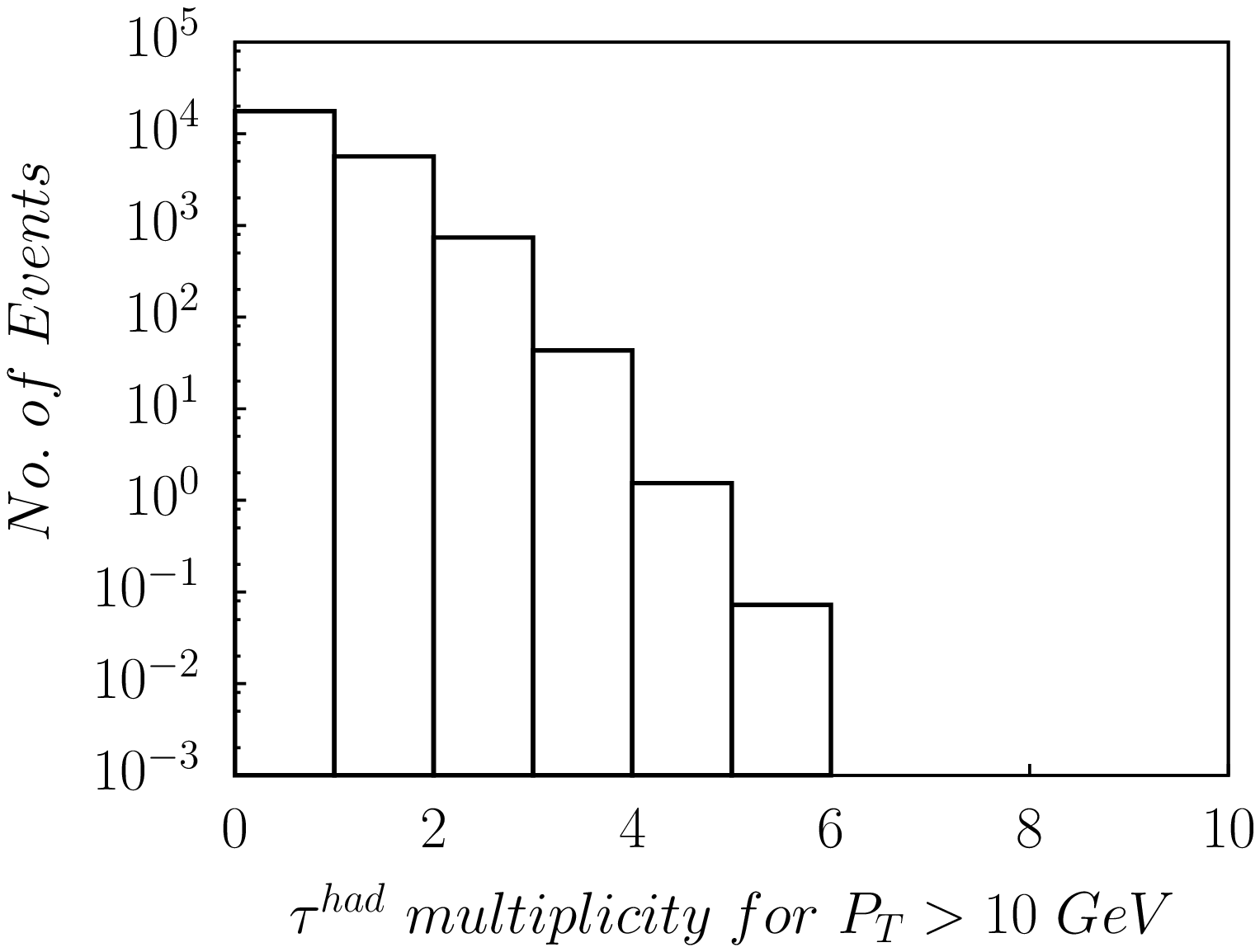,width=5.4cm,height=2.7cm}
\vspace*{-0.3cm}
\epsfig{file=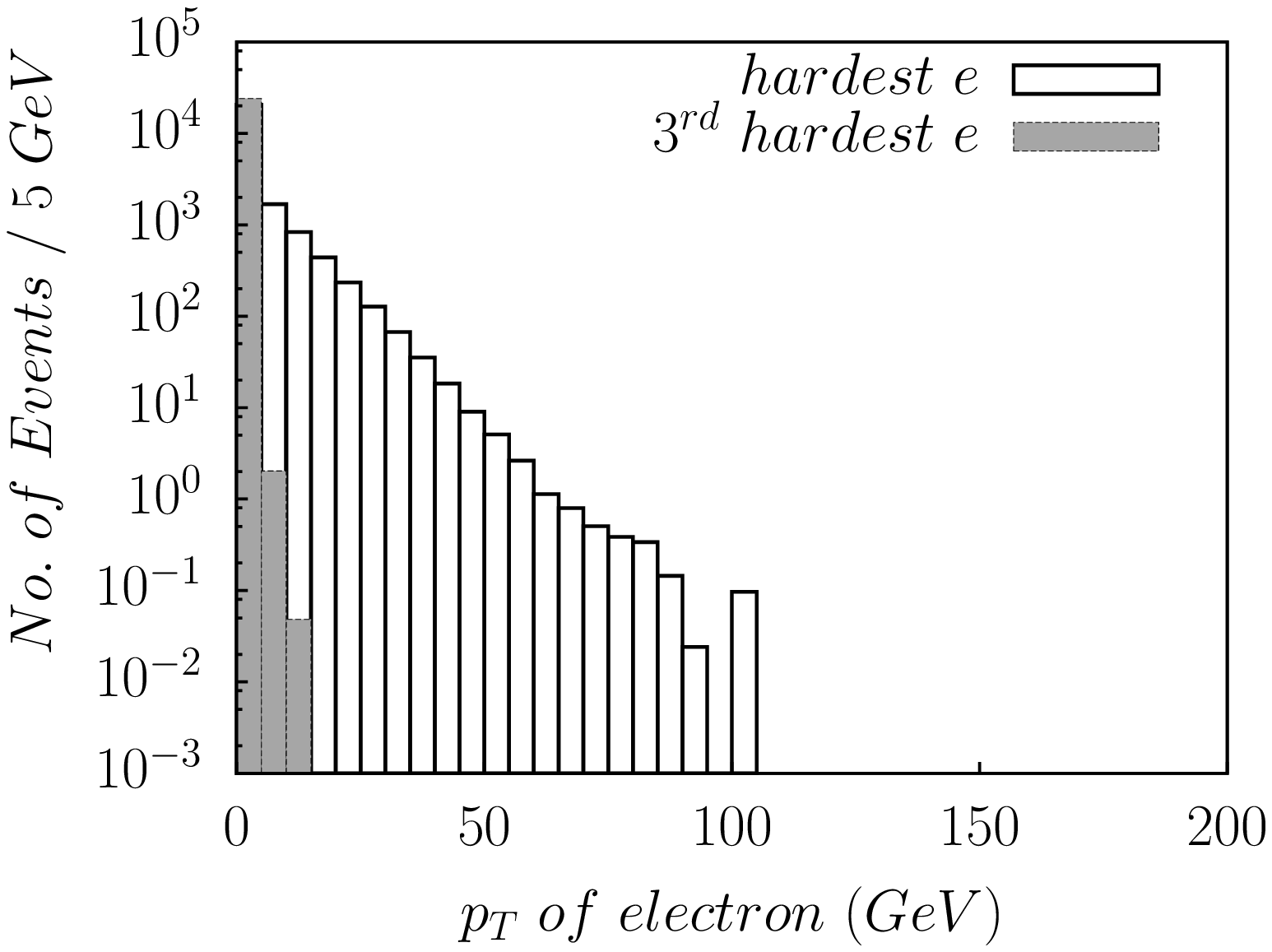,width=5.4cm,height=2.7cm} 
\epsfig{file=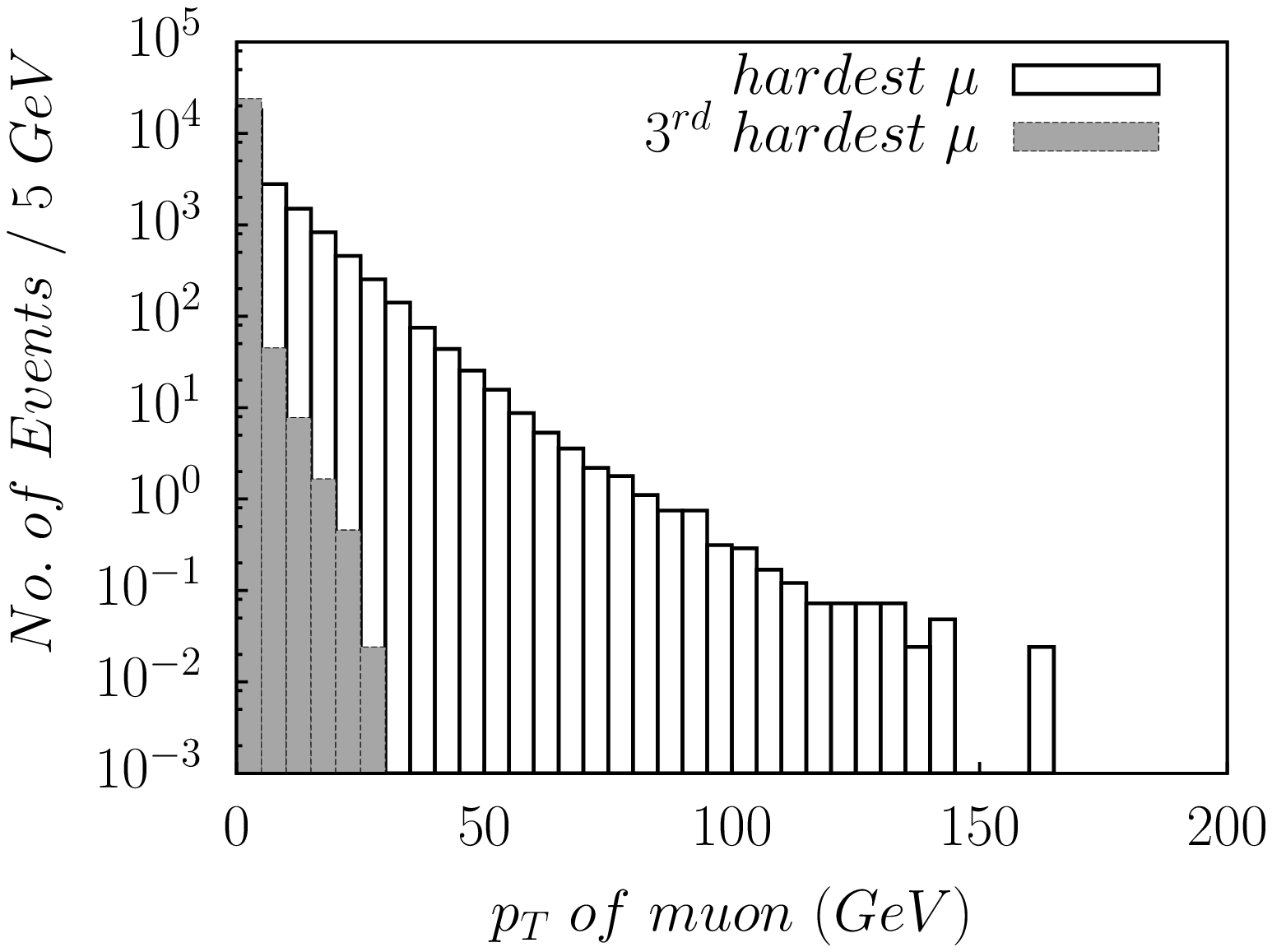,width=5.4cm,height=2.7cm} 
\epsfig{file=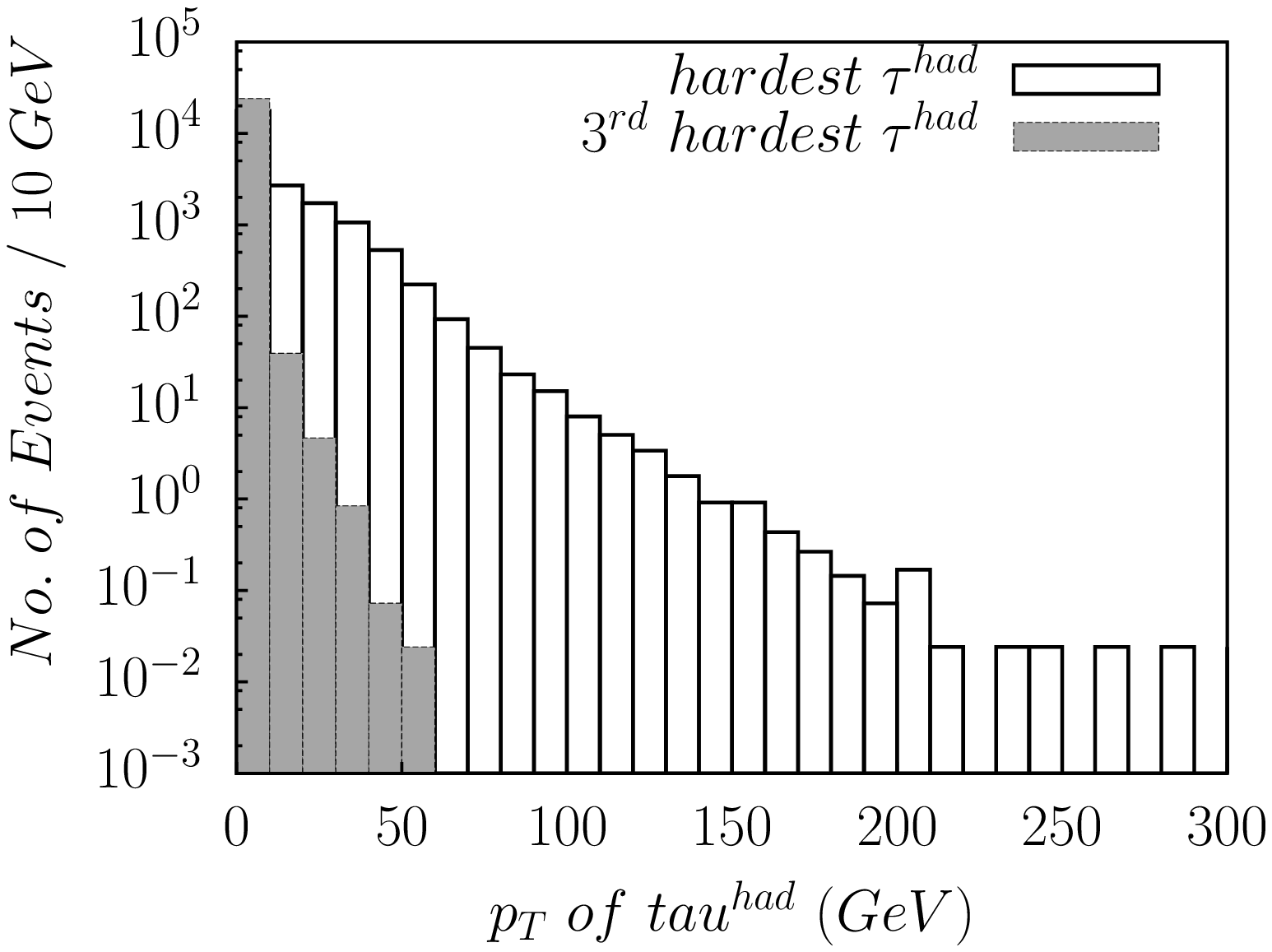,width=5.4cm,height=2.7cm} 
\caption{Multiplicity (top row) for $e$ (left), $\mu$ (middle) and 
hadronically decaying $\tau$ (right) with $p_{\rm T}>10$~GeV. $p_{\rm T}$ 
distributions (bottom row) for the leading (white) and the $3^\text{rd}$ 
leading (light grey) $e$ (left), $\mu$ (middle) and hadronically decaying 
$\tau$ (right). These plots correspond to $\sqrt{s}=8$ TeV with 
$\mathcal{L}=20$~fb$^{-1}$.}
\label{lepmul-leppt}
\end{figure*}

The values of the low-energy dimensionless free parameters that we assume are: 
$\lambda_i=0.11$, $\kappa_{111}=-0.0073 $, $\kappa_{222}=-0.0075 
$, $\kappa_{333}=-0.0077$, setting all other $\kappa_{ijk}$ to zero. 
$\kappa_{iii}$ are taken to be pseudo-degenerate for simplicity.
With universal $\kappa_{iii}$, only one linear combination of the right-handed 
neutrinos mix in an efficient way with the MSSM neutralinos. Then, depending 
on the sign of $\kappa$, $\neutralino_4$ or $\neutralino_{6}$ will
have a significant MSSM neutralino component. In our benchmark point where 
$m_{\neutralino_4} \approx 9.6 \gev$, $m_{\neutralino_5} \approx 11.5 \gev$, 
$m_{\neutralino_6} \approx 11.9 \gev$,
MSSM neutralino
admixture is sizable ($\sim $2\%) in $\neutralino_4$
and we get $Br(h_4 \to \neutralino_4 \neutralino_4)\approx 6\%$.
If $\kappa_{iii}>0$ were selected instead, then $h_4$ would decay
mainly to $\neutralino_6 \neutralino_6$, followed by fast decays 
$\neutralino_6 \to \neutralino_{4,5} \, \mu^+ \mu^-$, $\neutralino_5 
\to \neutralino_{4} \, \mu^+ \mu^-$. 
Since the produced muons are soft, they are 
difficult to trigger due to their very low transverse momentum, $p_{\rm T}$. 
Thus we stick to $\kappa_{iii}<0$, so that 
the relevant signal would be produced by the $h_4 \to 
\neutralino_4 \neutralino_4$ decay.
In the scalar sector relevant masses are: 
$m_{P_1}\approx3.6 \gev$, $m_{P_2}\approx3.8 \gev$, $m_{P_3}\approx5.5 \gev$,
and $m_{h_1}\approx 7.5 \gev$, $m_{h_2}\approx 8.0 \gev$, $m_{h_3}\approx 19.6 \gev$
with $m_{h_4}\approx 125.7 \gev$. Consequently, $\neutralino_4$
decays to $P_{i}+\nu$ is favourable with larger
available phase space. Although, two-body decays of
$\neutralino_4$ are kinematically possible, 
we have nevertheless computed the three-body decays for
greater accuracy. With chosen mass spectrum, 
$Br(\neutralino_4 \to \sum_{i=1}^3 \neutralino_i \tau^+ \tau^- ) \approx 99\%$, while 
remaining $1\%$ is shared between $\neutralino_{i} \mu^+ \mu^-$ and 
$\neutralino_{i} q_j \bar{q_j}$. Thus, the schematic $h_4$
decay chain studied is $h_4\to \neutralino_4 \neutralino_4 
\to 2 h^*_i/P^*_i 2 \nu \to 2\tau^+ 2\tau^- 2\nu$.

The matrix $Y_{\nu_{ij}}$ and the $\vevs$ of the left-handed sneutrinos, $\nu_i$, are 
connected to the reproduction of neutrino-physics. 
The processes $h_i/P_i\to \ell^+\ell^-$, on the other hand, are, to a 
very good approximation, independent of $Y_{\nu_{ij}}$ (and $\nu_i$). 
Hence, multiplicity of the charged leptons in the process 
is practically independent of $\nu_i$ and $Y_{\nu_{ij}}$.
As a corollary, a range for $Y_{\nu_{ij}}$ and $\nu_i$ 
predicting correct neutrino physics is well anticipated 
without drastic alteration in the event topology with
displaced multilepton.

For the studied benchmark point neutrino-sector parameters have been
chosen to result $m_{\neutralino_3} \sim 4.9 \times 10^{-11} \gev$,
giving a decay width $\Gamma_{\neutralino_4} \approx 6.7 \times 
10^{-16}\gev^{-1}$, which corresponds to a proper lifetime $\tau_{\neutralino_4}\approx 
10^{-9}$~s. One can avail the underlying relation between neutrino physics 
and $\wt\chi^0_4$ decay kinematics in the $\mu\nu$SSM through a 
common set of parameters $Y_{\nu}$, $\kappa$, $\lambda$, $\nu^c$, 
gaugino masses, etc.~\cite{Hirsch2009, Ghosh2010cu}, 
to obtain shorter $\tau_{\neutralino_4}$ through an increase
in absolute neutrino mass scale. Another viable handle
to modify $\tau_{\neutralino_4}$ is through a change
in $\neutralino_4$ composition, namely by altering relative dominance 
of gaugino masses, right-handed neutrino Majorana mass, $2 \kappa \nu^c$ 
and the $\mu$-parameter, $3\lambda\nu^c$~\cite{Ghosh2008, Ghosh2010cu, Hirsch2009}.
We have checked that variations in the 
benchmark point do not induce significant changes in the lepton multiplicity, but 
do affect $\Gamma_{\neutralino_4}$, \ie\ $\tau_{\neutralino_4}$, 
which remains in the experimentally accessible range of mm to m.

For numerical studies, {\tt PYTHIA (6.4.09)} \cite{Sjostrand:2006za} has been 
used as the MC event generator with default parton distribution function
at 8 TeV center-of-mass energy ($\sqrt{s}$).
The renormalisation/factorisation scale  is set equal to the parton-level 
$\sqrt{s}$. The initial and final state radiation and multiple 
interactions are kept switched on. The mass spectrum and decays are computed
with a custom-developed code. 
For the SM-like $h_4$, $gg\to h_4$ cross section ($6.51$~pb)
is rescaled by the reduced coupling~\cite{MuNuLHC}, 
yielding a next-to-next leading order $gg\to h_4$ 
production cross-section of $19.3$~pb~\cite{LHCwiki}.
{\tt PYTHIA} outputs are passed through {\tt PGS4}~\cite{PGS4} 
to simulate the detector response. All resulting 
distributions have been rescaled to an expected integrated luminosity of  
$\mathcal{L}=20$ fb$^{-1}$ corresponding to the full 2012 dataset.

The $\mu\nu$SSM is characterised by the production of several high-$p_{\rm T}$ 
leptons, as demonstrated in Fig.~\ref{lepmul-leppt} (top), where the $e$, $\mu$ 
and hadronically  decaying $\tau$ $(\tau^{had})$ multiplicity distributions are drawn 
for leptons with $p_{\rm T}>10\gev$. 
$e$ and $\mu$, are produced through the leptonic $\tau$ decays, although
$\mu$ pair can appear directly through ${h_i/P_i}$ decay.
With the chosen decay mode
the $\tau$ multiplicity is considerable even though the 
$\tau$-identification efficiency is much lower ($\sim50\%$) when compared to 
that of $e$ and $\mu$ ($\gtrsim95\%$). Occasionally highly collimated QCD 
jets can fake $\tau^{had}$'s and, as a result, $\tau$ multiplicity 
exceeds the expected number of $4$. This faking, however
disappears with a higher $p_{\rm T}$ cut. 

The $p_{\rm T}$ distributions of the leading and of the $3^\text{rd}$ 
leading lepton are shown in the bottom row of Fig.~\ref{lepmul-leppt}. It is evident 
that the leading lepton is energetic enough to trigger the event, should a 
single-lepton trigger is deployed. The rest of the leptons have sufficient 
$p_{\rm T}$ to be selected by a multilepton-based analysis, such as the ones 
developed by CMS~\cite{multileptons-CMS} and ATLAS~\cite{multileptons-ATLAS}. 
For instance, for the third leading $e$, $\mu$, and $\tau^{had}$, 
around 0.05, 10 and~43  events with $p_{\rm T}>10\gev$ are expected, respectively.
Clearly, multi-electron signature is least promising.

\begin{figure*}[htbp]
\centering 
\epsfig{file=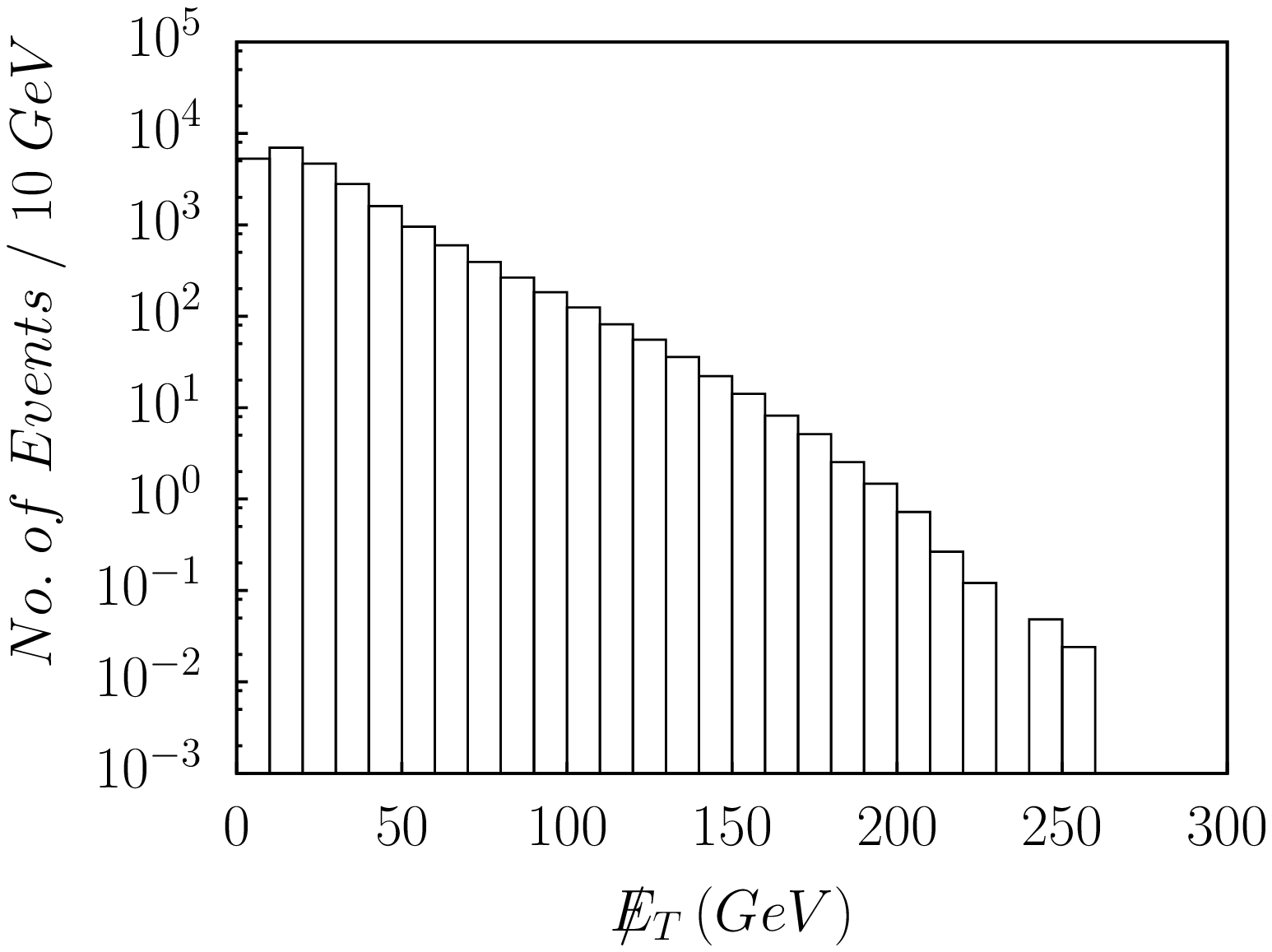,width=5.40cm,height=3.3cm}
\epsfig{file=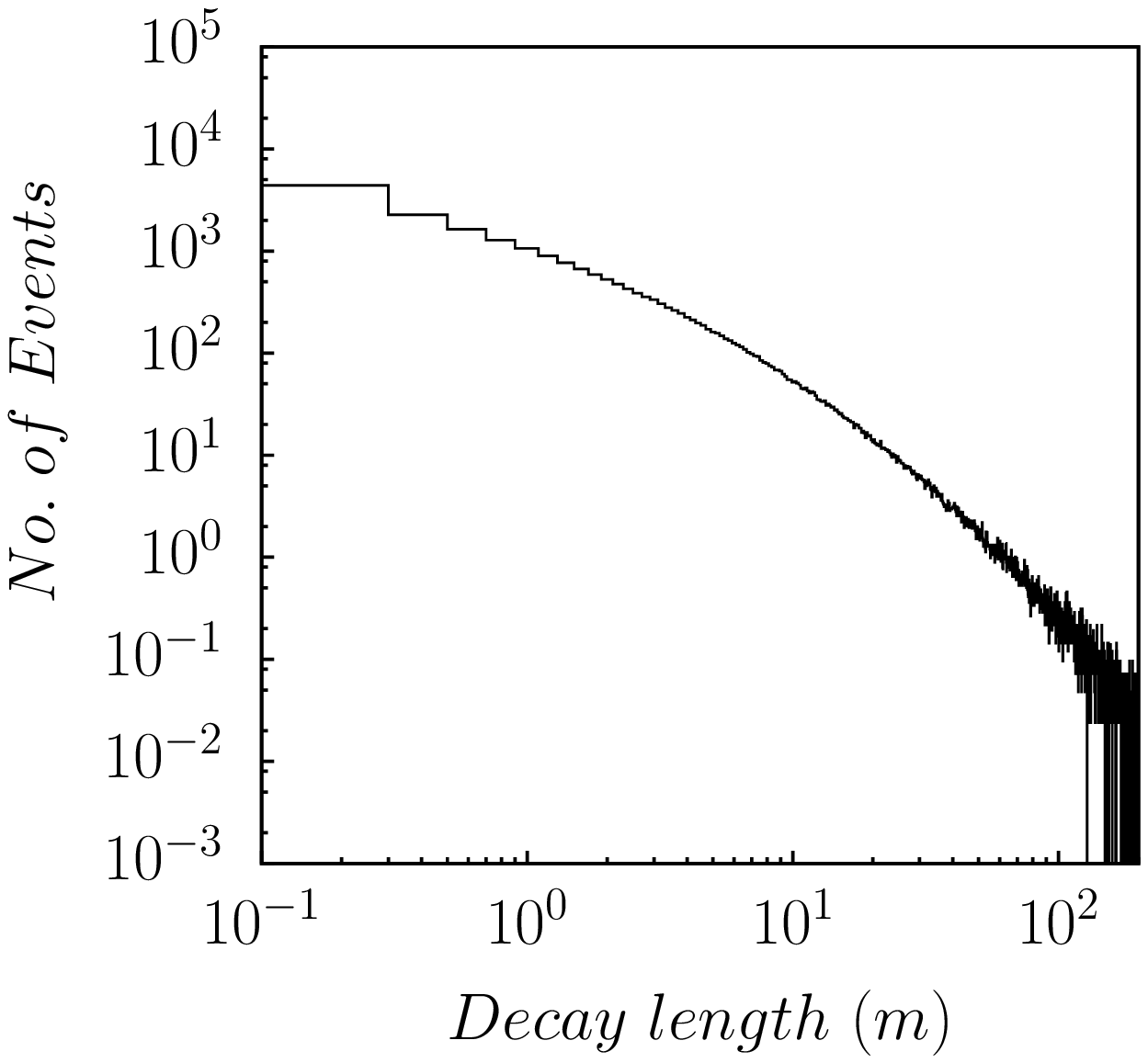,width=5.40cm,height=3.3cm}
\epsfig{file=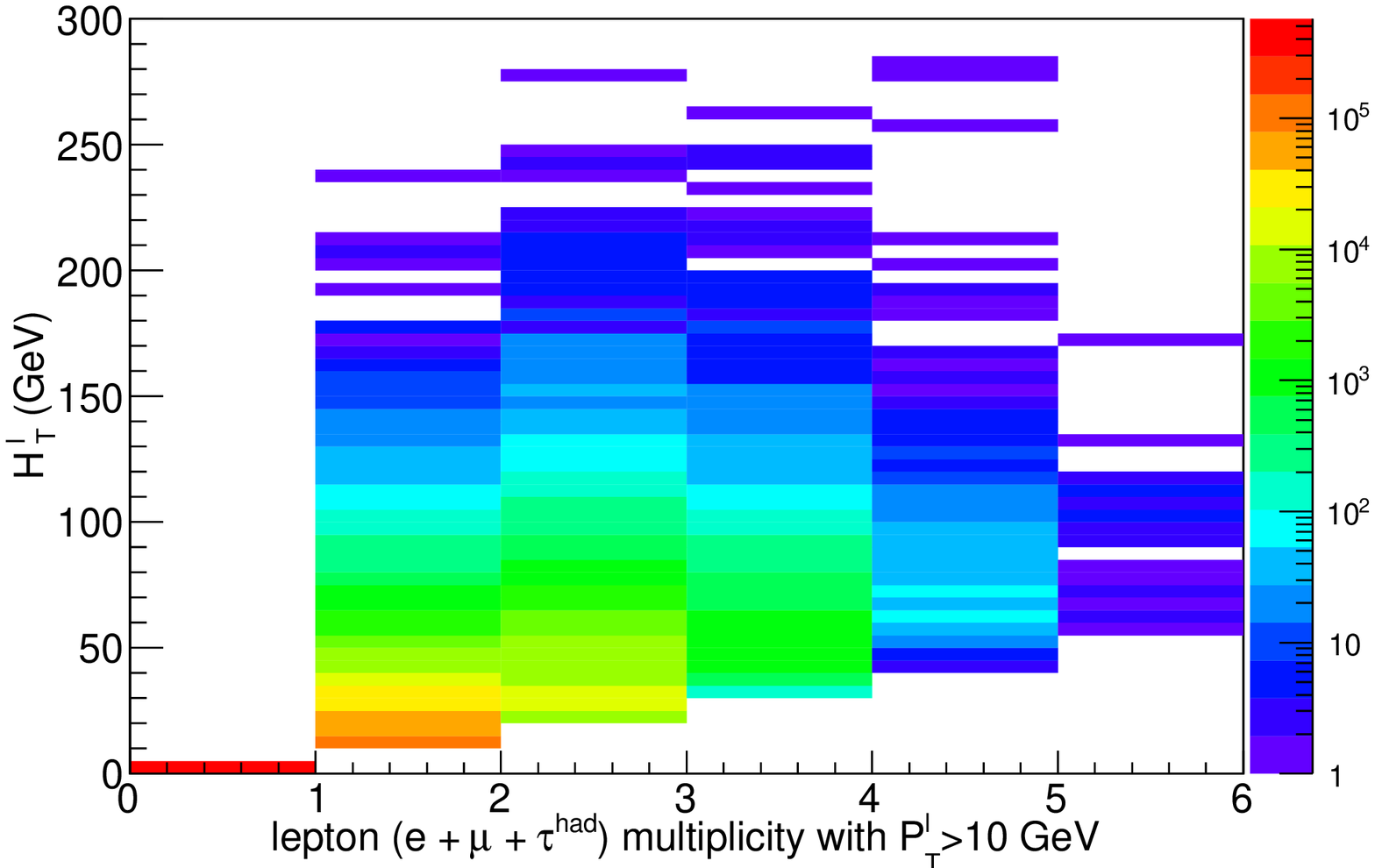,width=5.40cm,height=3.3cm}

\caption{$\MET~$~distribution (left),
$\neutralino_4$ decay-length distribution (middle)
 and $H{\rm _T^\ell}$ versus lepton multiplicity 
(right) for $\sqrt{s}=8 \tev$ and $\mathcal{L}=20 \fb^{-1}$.}
\label{met}
\end{figure*}
\begin{figure*}[htbp]
\vspace*{-0.50cm} 
\centering
\epsfig{file=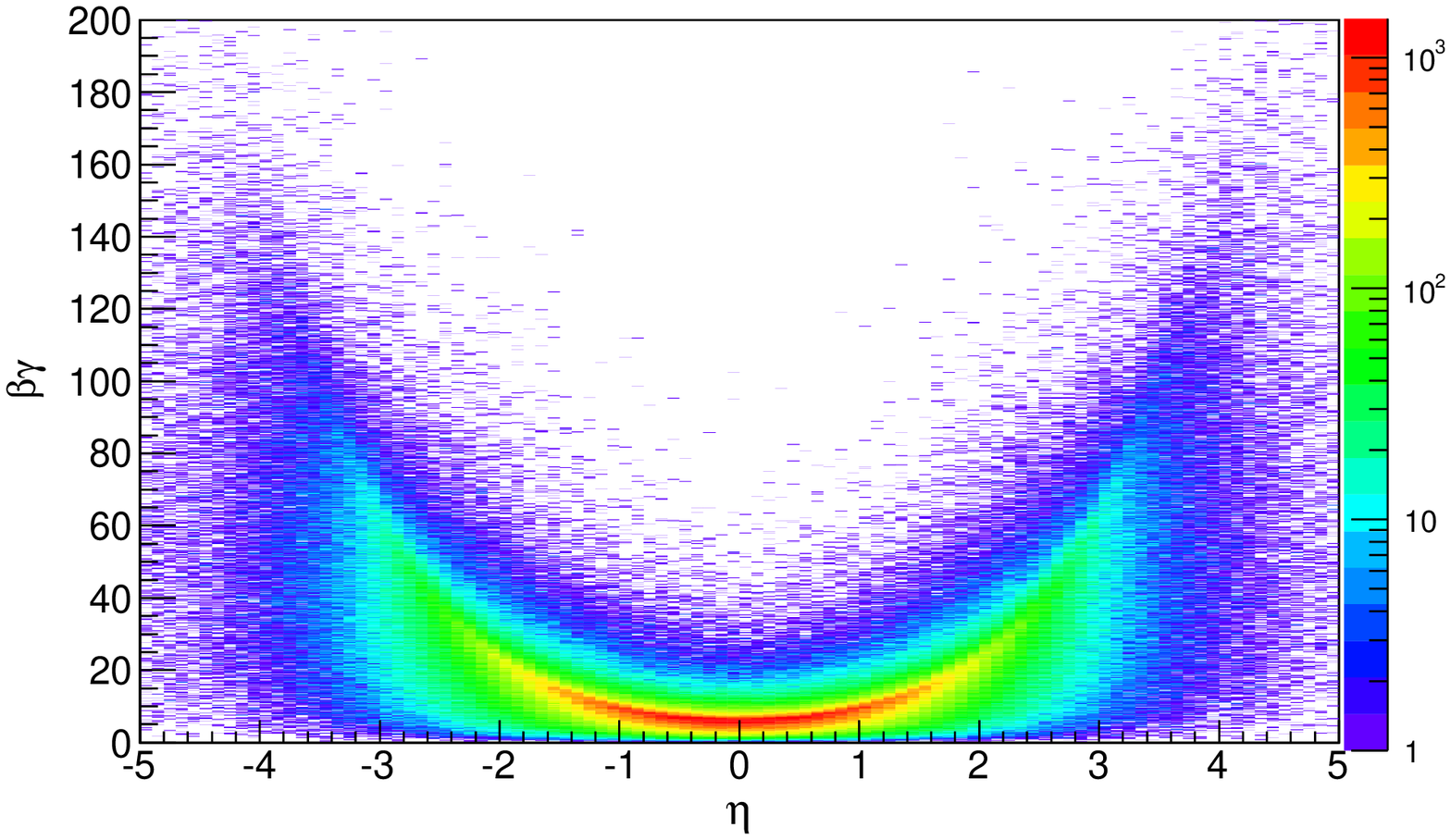,width=5.4cm,height=3.3cm}
\epsfig{file=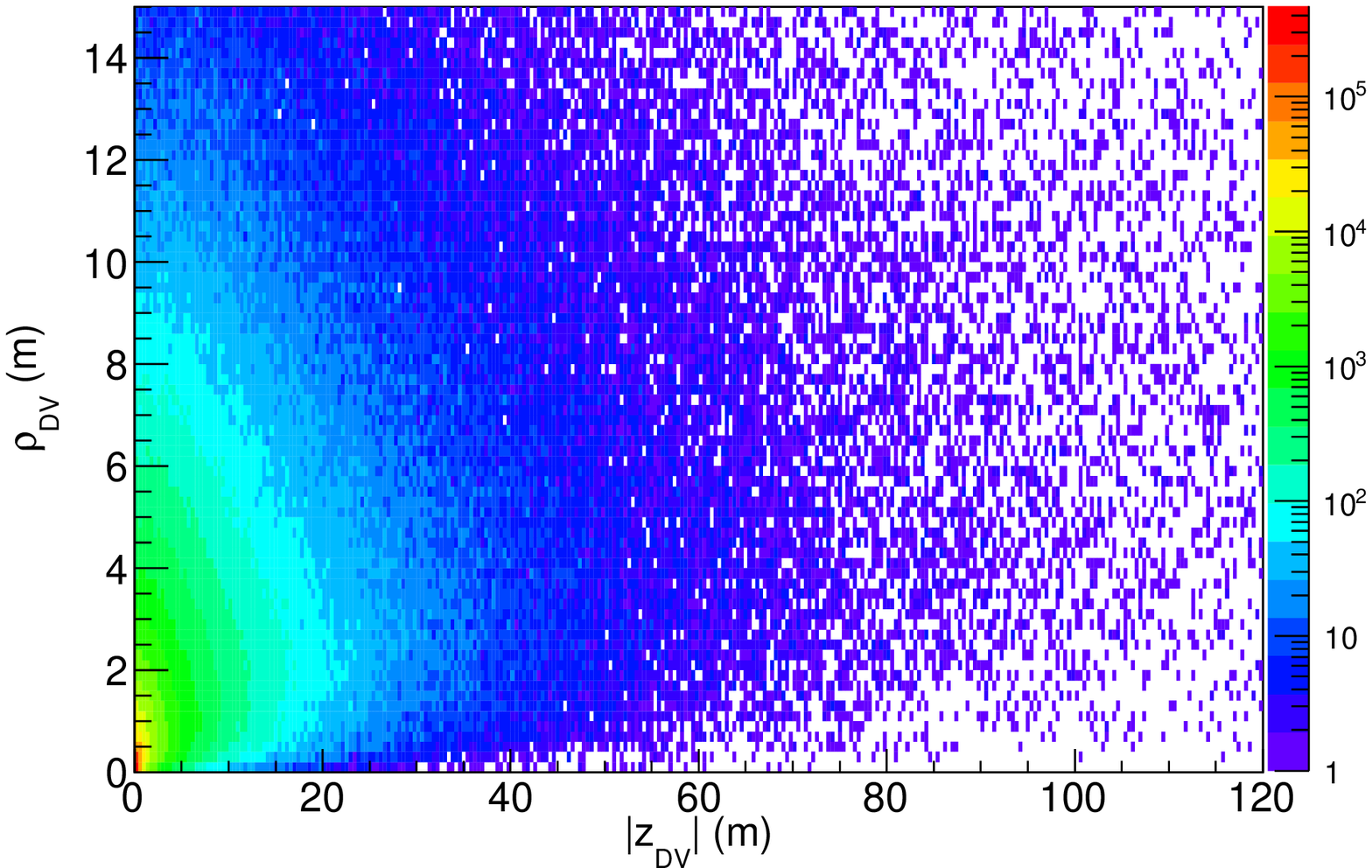,width=5.4cm,height=3.3cm}
\epsfig{file=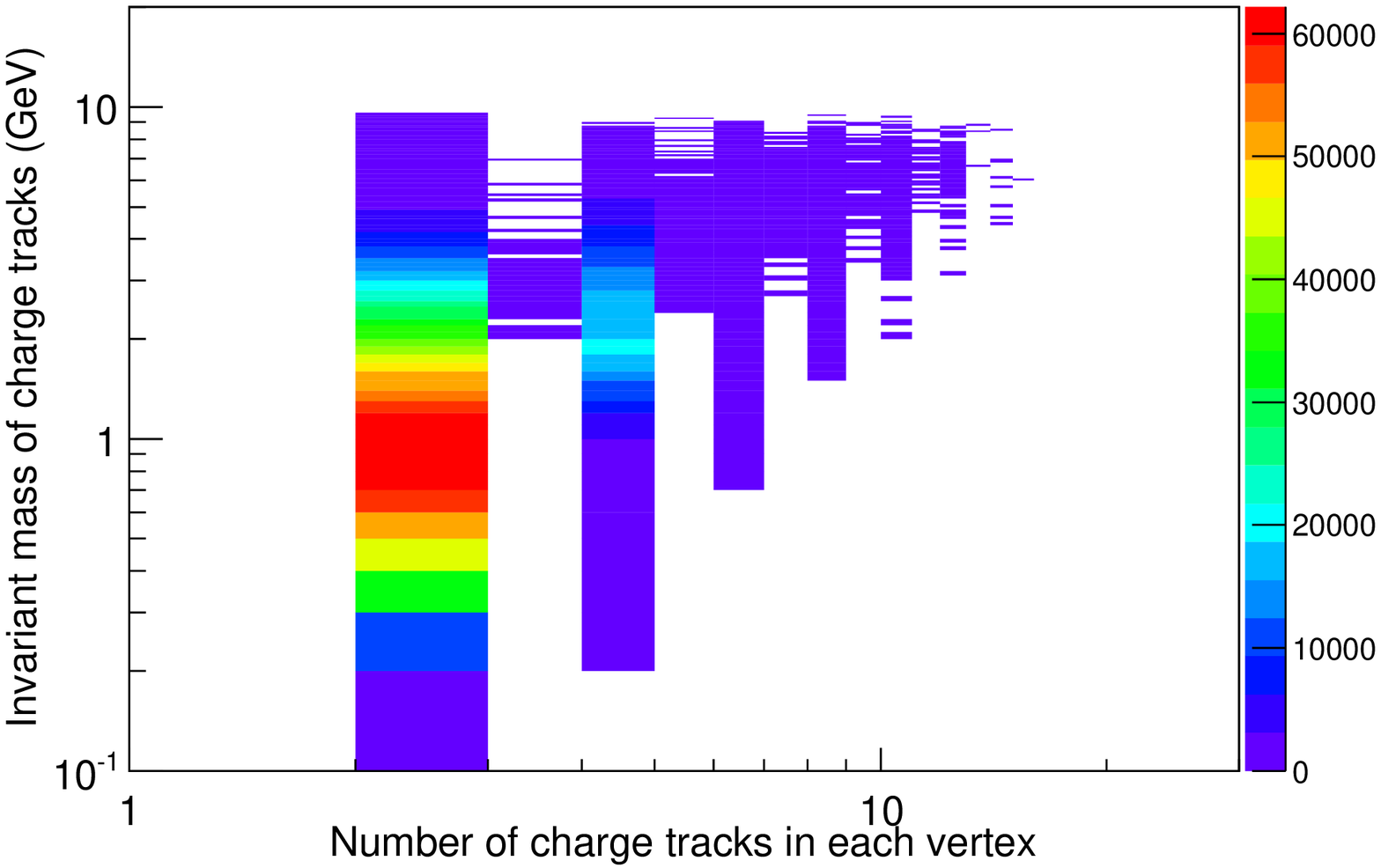,width=5.4cm,height=3.3cm}
\caption{$\beta\gamma$ versus $\eta$ (left), $\rho_{\rm DV}$ versus 
$|z_{\text{DV}}|$ (middle) and  charged-track mass versus the number of 
charge particles in each vertex (right) for  $\neutralino_4$ and for 
$\sqrt{s}=8\tev$ and $\mathcal{L}=20\fb^{-1}$.}
\label{collider-variables}
\end{figure*}

Such analyses, apart from the requirement of at least three or four leptons 
(including taus),  require a high value of \met\ or of the scalar sum of 
reconstructed objects: leptons, jets and/or \met. 
For the chosen signal many neutrinos ($\gsim 6$) appearing
in the final state from $\neutralino_4$ and from $\tau$ decay give 
rise to moderately
high missing transverse energy, \met, as depicted in Fig.~\ref{met} (left). Besides 
\met, the scalar sum of the $p_{\rm T}$ of all reconstructed leptons, $H{\rm _T^\ell}$, 
is also high in such events, as shown in Fig.~\ref{met} (right). Alternatively, the 
sum of \met\ and $H{\rm _T^\ell}$ can be deployed 
for further background rejection. These observables can provide additional 
handles when selecting events with many leptons. Also the invariant masses,
$m_{\ell^+\ell^-}$ and $m_{2\ell^+2\ell^-}$ are useful for 
the purpose of signal distinction.

A word of caution is due here. In the discussion on multilepton analyses so 
far, \emph{prompt} leptons are selected after imposing an upper limit
on the transverse and the longitudinal impact parameters, in order to reject cosmic-ray 
muons and insure good-quality track selection. Such a selection criterion should 
be relaxed or even reversed, if sensitivity to the $\mu\nu$SSM events is sought 
after. The reason stems from the long lifetime of the $\neutralino_4$ and hence 
the DV that its decay creates. This feature is quantified in the 
middle plot of Fig.~\ref{met}, were the decay-length distribution is drawn. 
As expected from the proper decay length of $c\tau_{\neutralino_4}\approx 30$~cm, 
in a significant percentage of events, $\neutralino_4$ decays inside the tracker, 
\eg\ 28\% of events decay within $30 \cm$ and $44$\% events within 1~m. Therefore, 
the $\mu\nu$SSM signal events will be characterised by displaced $\tau$-leptons 
plus neutrinos. This distinctive signature opens up the possibility to exploit 
current or future variations of analyses carried out by ATLAS and CMS looking for 
a displaced muon and tracks~\cite{DV-ATLAS} or searching for displaced 
dileptons~\cite{DL-CMS} or  muon jets~\cite{DL-ATLAS} arising in Higgs decays 
to pairs of long-lived invisible particles.

The kinematics of the DVs and their products are demonstrated in 
Fig.~\ref{collider-variables}. The $\neutralino_4$ boost, expressed by 
$\beta\gamma$, where $\beta$ is $\neutralino_4$ velocity over $c$ and $\gamma$ 
the Lorentz factor, versus the pseudorapidity $\eta$ is shown on the left. 
The shape reflects the fact that a single particle ($h_4$) is produced at the 
hard scattering of $pp$ collisions, hence low momentum is expected in the 
central region. The average boost is comparable to the signal analysed in 
an ATLAS search for a muon and tracks originating from DVs~\cite{DV-ATLAS}. 
The boost affects the efficiency with which such a DV can be reconstructed, 
since high $\neutralino_4$ boost leads to collimated tracks difficult to 
differentiate from primary vertices.

In the middle plot in Fig.~\ref{collider-variables}, the spacial distribution 
of a DV is displayed in cylindrical coordinates. A large fraction of DVs falls 
in the inner-tracker volume of an LHC experiment, \ie\ $\rho_{\text{DV}}\lesssim 
1$~m and $|z_{\text{DV}}|\lesssim 2.5$~m, thus DVs arising in the $\mu\nu$SSM 
should be detectable at LHC, either with existing analyses~\cite{DV-ATLAS,DL-CMS,DL-ATLAS} 
or via variations of those to search for displaced taus and \met.

In the right of Fig.~\ref{collider-variables}, we show the correlation 
between the number of charged tracks in each DV, $N_{\rm trk}$, and their 
invariant mass, $m_{\rm DV}$. A selection of high-$N_{\rm trk}$ and high-$m_{\rm DV}$, 
has been demonstrated~\cite{DV-ATLAS} to efficiently 
suppress background from long-lived SM particles ($B$-mesons, kaons). The modulation 
observed in $N_{\rm trk}$ is due to the one-prong or three-prong hadronic $\tau$ decays. 

Summarizing, the $\mu\nu$SSM could be tested at the LHC through the 
production of a Higgs-like scalar with a mass about $125\gev$ which decays into a 
pair of light long-lived neutralinos. Such events could be probed by 
ATLAS and CMS with the currently available $8\tev$ data in two ways: either by 
looking for multilepton events produced in the SUSY cascade decay chain, when 
relaxing or even reversing the requirement for the leptons to come from 
the primary vertex, or by searching for tracks not-pointing back to the 
primary vertex, originating from a secondary vertex. 
In either case, a moderately high missing 
transverse energy due to multiple 
neutrinos is expected. In principle, other Higgs 
boson decay chains or other processes might have been addressed to test the 
$\mu\nu$SSM. However, we leave 
this necessary task for a future work \cite{future}.

PG thanks S.~Biswas, K.~Ghosh and C.~B.~Park  for  insightful discussions.
The work  of PG and CM was supported in part by the Spanish MINECO under 
grants FPA2009-08958 and FPA2012-34694, and under
the `Centro de Excelencia Severo Ochoa' Programme SEV-2012-0249,
by the Comunidad de Madrid under grant HEPHACOS S2009/ESP-1473, and by the
European Union under the Marie Curie-ITN program PITN-GA-2009-237920.
The work of DL was supported by the Argentinian CONICET. The work of VM was 
supported by the Spanish MINECO under grant FPA2009-13234-C04-01 and by the 
Spanish AECID under PCI project A1/035250/11. The work of RR was supported 
by the Ram\'on y Cajal program of the Spanish MINECO
and also thanks the support of the MINECO under grant FPA2011-29678.
The authors also acknowledge the support of the MINECO's Consolider-Ingenio 2010 Programme under grant 
MultiDark CSD2009-00064.



\end{document}